\documentclass{jps-cp}
\usepackage{txfonts} %Please comment out this line unless the txfonts package is availabe in your LaTeX system.
\usepackage{bm}

\title{Strangeness Enhancement
in p+p, p+Pb and Pb+Pb Collisions at LHC Energies}

\author{Yuuka \textsc{Kanakubo}$^{1}$, Michito \textsc{Okai}$^{1}$, Yasuki \textsc{Tachibana}$^{2}$ and Tetsufumi \textsc{Hirano}$^{1}$}

\inst{$^{1}$Department of Physics, Sophia University, 7-1 Kioi-cho, Chiyoda-ku, Tokyo 102-8554, Japan\\
$^{2}$
Department of Physics and Astronomy, Wayne State University, 666 W. Hancock St., Detroit, MI 48201, USA
}
\email{
y-kanakubo-75t@eagle.sophia.ac.jp
}
\recdate{\today}

\abst{
Recently ALICE Collaboration reported 
enhancement of yield ratio of multi-strange hadrons to charged pions as a function of multiplicity at mid-rapidity
in proton--proton (p+p), proton--lead (p+Pb) and lead--lead (Pb+Pb) collisions at the LHC. 
Motivated by these results, we have developed the dynamical core--corona initialization framework
which enables us to describe p+p and p+Pb collisions as well as Pb+Pb, and we investigate whether the quark gluon plasma (QGP) is
created in small colliding systems by analyzing various hadron yields and their ratios systematically. 
We find that our results reproduce tendencies of the ALICE data especially for multi-strange hadrons. 
These results indicate that the QGP is partly formed in high multiplicity events in small colliding systems.
}

\kword{Quark gluon plasma, Relativistic heavy-ion collisions, Small colliding systems, Core--corona picture, Strangeness enhancement}
\begin{document}
\maketitle

\section{Introduction}
High energy heavy-ion collision experiments are performed 
at the Relativistic Heavy Ion Collider (RHIC) in Brookhaven National Laboratory
and the Large Hadron Collider (LHC) in CERN
to understand properties of the quark gluon plasma (QGP).
It is known that various experimental data are described by relativistic hydrodynamics, 
which indicates that the QGP behaves nearly like a perfect fluid.
Conventionally, it has been assumed that the QGP is generated only in heavy-ion collisions and that small systems such as proton–-proton or proton--nucleus collisions provide references for extracting medium effects in heavy-ion collisions.
Recently, ALICE Collaboration obtained surprising results which indicate, however, the QGP formation in small colliding systems \cite{ALICEpp}.
They measured yield ratios of multi-strange hadrons to charged pions
as functions of multiplicity at mid-rapidity and 
the results exhibit rapid increase with multiplicity in proton--proton (p+p) collisions.
Moreover, the ratios do not seem to depend on the system size or collision
energies. One of the possible description to interpret this result is the core--corona picture \cite{Werner, Aichelin, Becattini,Pierog,Bozek}.
The core--corona picture is a two-component description which is described by chemically equilibrated matter and unscathed partons.
In this study we introduce the core--corona picture into the dynamical initialization model which was proposed in Ref. \cite{Okai}.
Under the core--corona picture, initially produced partons tend to become fluids in dense region in which a lot of interactions among partons are assumed to happen, while partons do not tend to become fluids in dilute region in which few interactions occur.
We introduce the above picture into the fluidization rate in the dynamical initialization model \cite{Kanakubo} and analyze the multiplicity dependence of particle yield ratios in various colliding systems.
%We first siFmulate p+p, p+Pb and Pb+Pb collisions to compare with the experimental data.

%You can use this file as a template to prepare your manuscript for JPS Conference Proceedings\cite{cp,jpsj,instructions}.
%Copy \verb|jps-cp.cls| and \verb|cite.sty| onto an arbitrary directory under the texmf tree, for example, \verb|$texmf/tex/latex/jpsj|. If you have already obtained \verb|cite.sty|, you do not need to copy it.
%Many useful commands for equations are available because \verb|jps-cp.cls| automatically loads the \verb|amsmath| package. Please refer to reference books on \LaTeX\ for details on the \verb|amsmath| package.

\section{Model}
Firstly, we mention our model flow briefly. 
In this framework the QGP fluids are generated from partons produced initially just after the first contact of collisions.
First we generate the initial partons using a Monte Carlo event generator PYTHIA ver 8.230 \cite{Sjostrand, Bierlich},
with the hadronization option switched off in order to define phase space distributions of the partons.
Next we perform the dynamical core--corona initialization to obtain the initial condition of the QGP fluids. 
After the initialization, we solve ideal hydrodynamic equations with source terms
%,  $\partial_\mu T^{\mu\nu}=J^\nu$, 
in fully (3+1)-dimensional space as usual.
We start initialization of the QGP fluids from $\tau_{00}=0.1\ \rm{fm}$ which is assumed to be formation time of the partons and continue until $\tau_0 = 0.6 \ \rm{fm}$ which is initial time of the fluids. 
After the hydrodynamic simulations, we calculate final hadron yields from the core and the corona separately. 
We obtain final hadron yields from the core integrating the Cooper--Frye formula \cite{CF} at chemical freeze-out surface.
We consider the corrections of yields from resonance decays by multiplying 
factors estimated from a statistical model \cite{Andronic}.
%by the fractions of yields which is obtained in statistical model \cite{Andronic}. 
On the other hand, we calculate final hadron yields 
from the corona performing string fragmentation using PYTHIA.
Thus, the final hadron yield in this framework is the sum of these two final yields from both the core and the corona.

\subsection*{}
The source term in the hydrodynamic equation can be defined as
\begin{eqnarray}
\label{eq:source_j}
J^\mu(x)  =  -\sum_{i}  \frac{dp_i^\mu}{dt} G(\bm{x}-\bm{x}_i(t)),
\end{eqnarray}
where $p_i^\mu$ is the four momentum of the $i$ th parton obtained from PYTHIA and the summation is taken over all partons in the event.
%We solve the hydrodynamic equations with source terms denoted as $\partial_\mu T^{\mu\nu}=J^\nu$ and
%we define it as eq.(\ref{eq:source_j}) where $p_i^\mu$ is the four momentum of the $i$-th parton obtained by PYTHIA. Summation is taken over all partons in the event.
Here we employ the Gaussian function, $G$, for smearing 
energy and momentum deposited at the position of the parton.
To take account of the core--corona picture, we parametrize the rate of energy and momentum deposition of the parton as
\begin{eqnarray}
\label{eq:parton-loss-rate}
\frac{d p_i^\mu}{dt}(t) &=& -a_0  \frac{\rho_i (\bm{x}_i(t))}{{p_{T, i}}^2(t)} p_i^\mu (t) , \\
%\hspace{25pt}
\rho_i(\bm{x}) &=& \sum_{j\neq i} G(\bm{x} - \bm{x}_j(t)).  
\end{eqnarray}
Here, $a_0$, $\rho_i$ and $p_{T,i}$ are control parameter for magnitude, density of partons surrounding the $i$ th parton and transverse momentum.
Under this formulation, partons in dense region are likely to become fluids while those in dilute region tend to survive.

\section{Results}
\begin{figure}[tbh]
\vspace{-0.7cm}
\begin{center}
\includegraphics[bb=0 0 454 520,width=0.40\textwidth]{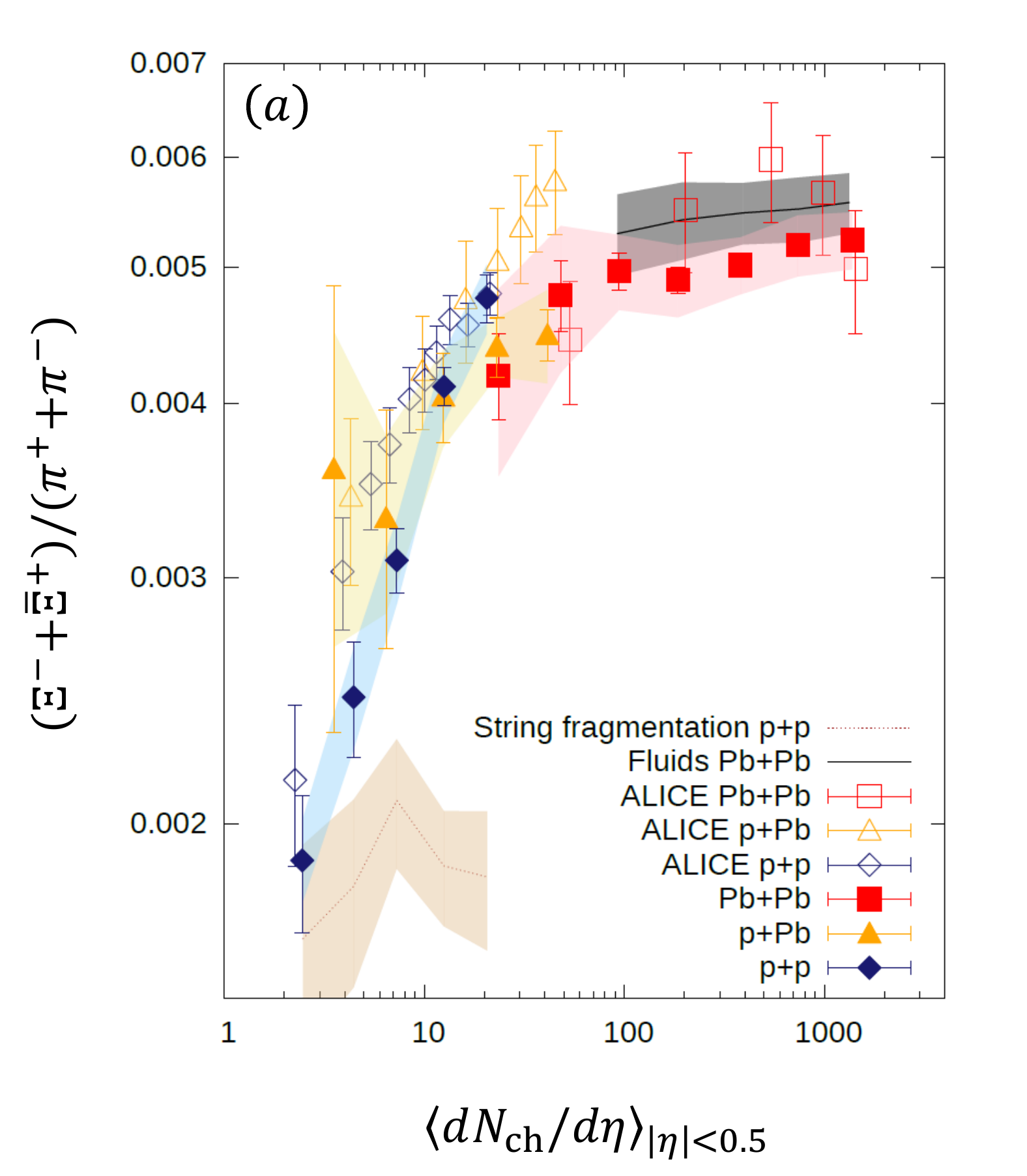}
\hspace{40pt}
\includegraphics[bb=0 0 454 520,width=0.40\textwidth]{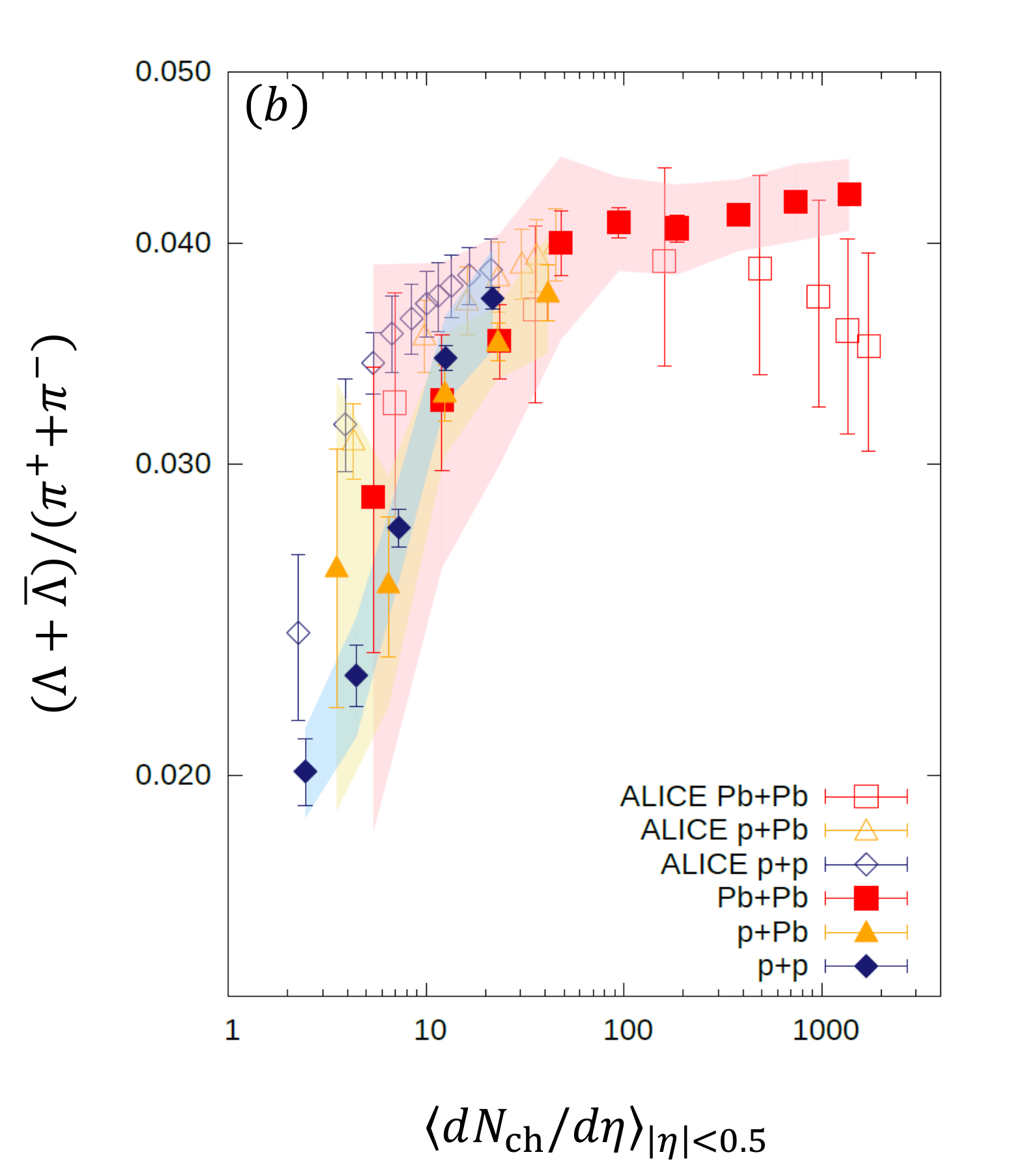}
\end{center}
%\vspace{5pt}
\begin{center}
\includegraphics[bb=0 0 454 520,width=0.40\textwidth]{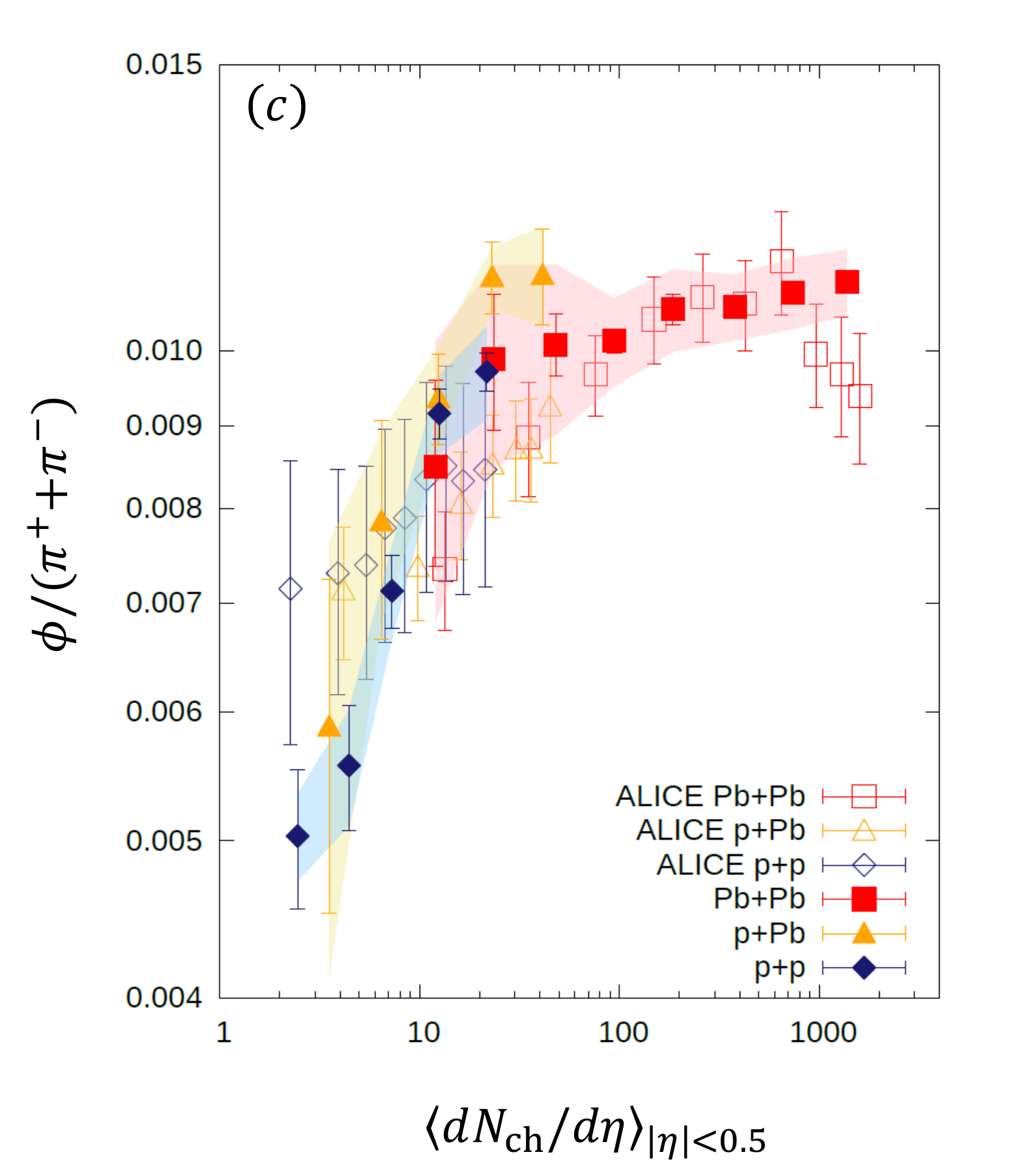}
\hspace{40pt}
\includegraphics[bb=0 0 454 520,width=0.40\textwidth]{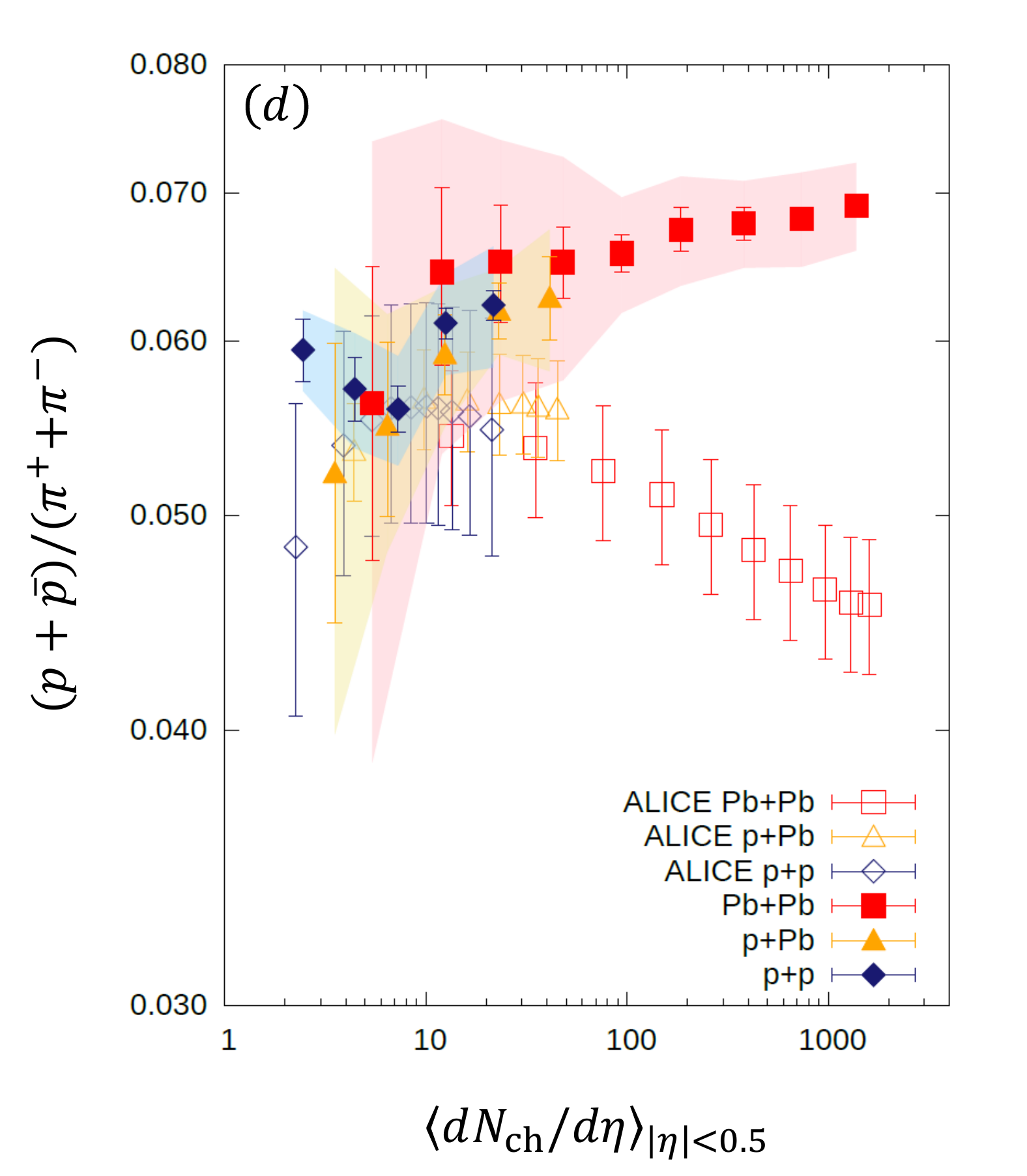}
\end{center}
\caption{
The particle yield ratio as a function of multiplicity in p+p (diamond), p+Pb (triangle) and Pb+Pb (square) at LHC energies. Closed symbols are our results while open symbols are ALICE experimental data \cite{ALICEpp, ALICE2, ALICE3, ALICE4, ALICE5, ALICE6, ALICE7}. Yield ratios of  (a) cascades, (b) lambdas, (c) phi mesons and (d) protons to charged pions are shown.
}
\label{f1}
\end{figure}

Collision systems and energies in these simulations are p+p, p+Pb and Pb+Pb collisions at $\sqrt{s_{\rm{NN}}}=7$, $5.02$ and $2.76$ TeV, respectively, and Au+Au collisions at $\sqrt{s_{\rm{NN}}}=200$ GeV.
Figure \ref{f1} shows the particle yield ratio in those collision systems as a function of multiplicity from our framework
compared with the experimental data \cite{ALICEpp, ALICE2, ALICE3, ALICE4, ALICE5, ALICE6, ALICE7}. 
For (a) cascades (${\Xi}^- + \bar{\Xi}^+$), (b) lambdas ($\Lambda + \bar{\Lambda}$) and (c) phi mesons ($\phi$), our results show good agreement with the experimental data.
The ratios increase up to $\langle dN_{\rm{ch}}/d\eta \rangle\sim100$ and saturate in high multiplicity events.
In Fig. 1(a), we also plot the yield ratios from string fragmentation and that from fluids as references.
Since our result in low multiplicity p+p events is almost identical with the one from string fragmentation, contribution from string fragmentation turns out to be dominant.
On the other hand, the yield ratio of full calculation increases with multiplicity towards the one from fluids. This tendency implies that 
the contribution of fluids becomes larger and dominant at high multiplicity events.
Moreover, our result seems to behave just as a function of multiplicity, and are independent of their system size or collision energies.  
For (d) proton ($p + \bar{p}$) yield ratio, a deviation between our result and the experimental data is seen above $\langle dN_{\rm{ch}}/d\eta \rangle \sim 50$.
This would be because proton and anti-proton annihilation could happen in the late hadronic rescattering stage.
We leave consideration of this effect for the future work.
\vspace{-0.7cm}

\begin{figure}
%\vspace{-2cm}
\begin{center}
\includegraphics[bb=0 0 340 520,width=0.37\textwidth]{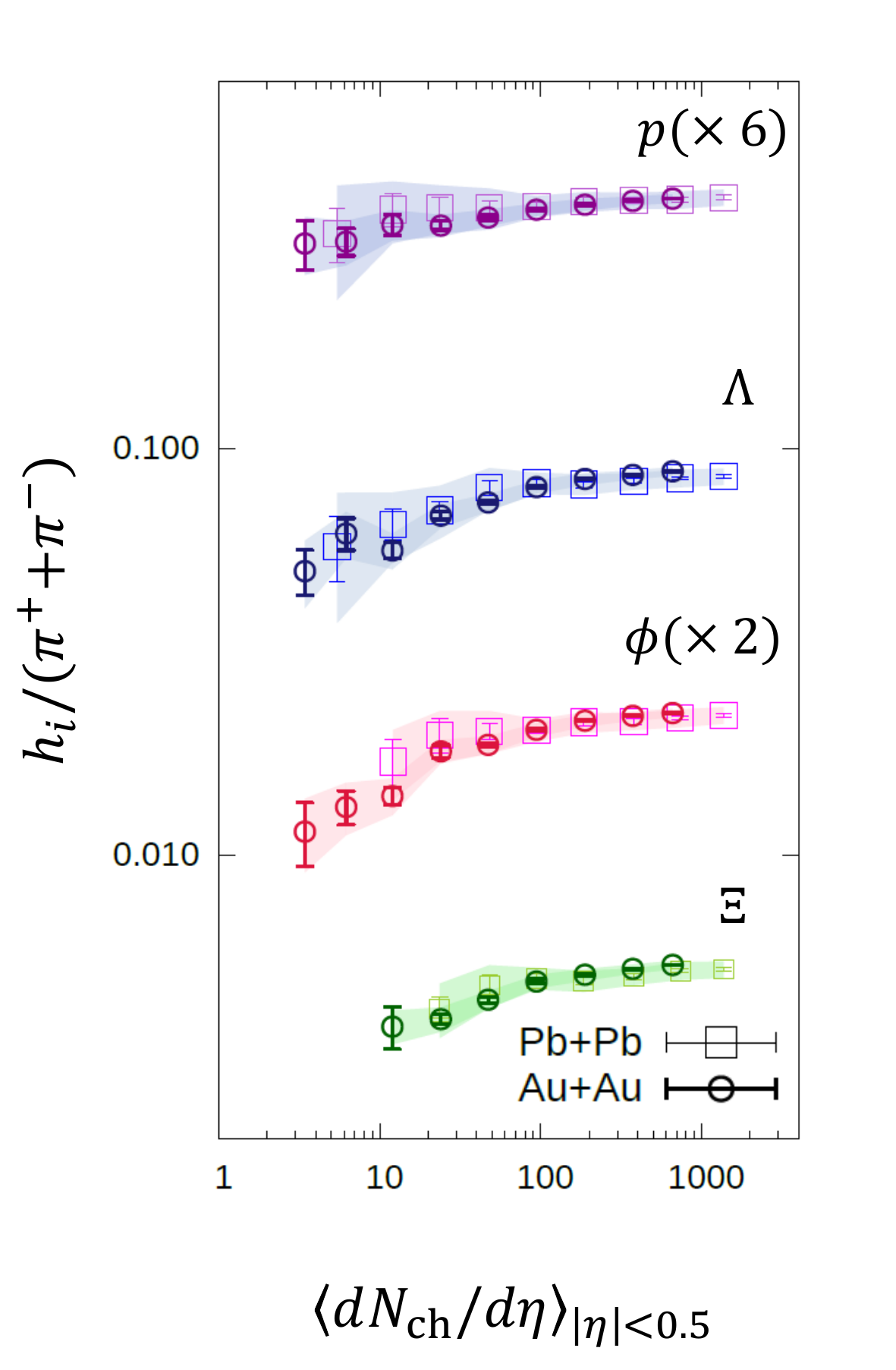}
\end{center}
\caption{Collision energy dependence of particle yield ratios as functions of multiplicity.
The comparison between
results in Pb+Pb collisions at the LHC energy and the ones in Au+Au collisions at the RHIC energy}
\label{f2}
\end{figure}

Figure \ref{f2} shows the collision energy dependence of particle yield ratios in heavy-ion collisions.
The yield ratios in Au+Au collisions scale with multiplicity as the ones in Pb+Pb collisions do, which demonstrates collision energy independence in yield ratios within this collision energy range.

\section{Summary}
In this study we introduce the core--corona picture into the dynamical initialization model and simulate high-energy nuclear collisions from small to large colliding systems.
The picture changes continuously from unscathed partons to QGP fluids with increasing multiplicity.
Our results show reasonable agreement with the ALICE experimental data in the yield ratios of cascades, lambdas and phi mesons to charged pions. 
These results demonstrate that there is no system size or collision energy dependences and that the ratio scales solely with multiplicity.
This result provides the strong indication of partial QGP generation in high multiplicity small colliding systems.

%Label figures, tables, and equations appropriately using the \verb|\label| command, and use the \verb|\ref| command to cite them in the text as ``\verb|as shown in Fig. \ref{f1}|". This automatically labels the numbers in numerical order.

%日本物理学会誌投稿規定によると \textit{et al}.　はイタリックじゃなくてもよい。
 

\begin{thebibliography}{17}
\bibitem{ALICEpp} J.~Adam \textit{et al}. (ALICE): Nature Phys., \textbf{13}, 535 (2017). 
\bibitem{Bozek} P.~Bozek: Acta Phys. Polon., \textbf{B36}, 3071 (2005).
\bibitem{Werner} K.~Werner: Phys. Rev. Lett., \textbf{98}, 152301 (2007).
\bibitem{Becattini} F. Becattini and J. Manninen: Phys. Lett., \textbf{B673}, 19 (2009).
\bibitem{Aichelin} J. Aichelin and K. Werner: Phys. Rev., \textbf{C81}, 029902 (2010).
%\bibitem{Steinheimer} J. Steinheimer and M. Bleicher, Phys. Rev., C84, 024905 (2011), arXiv:1104.3981.
\bibitem{Pierog} T. Pierog, Iu. Karpenko, J. M. Katzy, E. Yatsenko and K. Werner: Phys. Rev., \textbf{C92(3)}, 034906 (2015).
%\bibitem{Petrovici} M. Petrovici, I. Berceanu, A. Pop, M. Tarzila, and C. Andrei, Phys. Rev., C96(1), 014908 (2017),arXiv:1703.05805.
%\bibitem{Werner_2} K. Werner, A. G. Knospe, C. Markert, B. Guiot, Iu. Karpenko, T. Pierog, G. Sophys, M. Stefaniak,
%\bibitem{Bleicher} M. Bleicher, and J. Steinheimer, EPJ Web Conf., 171, 09002 (2018).
%\bibitem{Akamatsu} Y. Akamatsu, M. Asakawa, T. Hirano, M. Kitazawa, K. Morita, K. Murase, Y. Nara, C. Nonaka, and A. Ohnishi, Phys. Rev., C98(2), 024909 (2018), arXiv:1805.09024.
%\bibitem{Bozek_2} Piotr Bozek, Phys. Rev., C79, 054901 (2009), arXiv:0811.1918.
\bibitem{Okai} M.~Okai, K.~Kawaguchi, Y.~Tachibana and T.~Hirano: Phys. Rev., \textbf{C95(5)}, 054914 (2017).
\bibitem{Kanakubo} Y.~Kanakubo, M.~Okai, Y.~Tachibana and T.~Hirano: PTEP, \textbf{2018 no.12}, 121D01 (2018).
\bibitem{Sjostrand} T.~Sjostrand, S.~Mrenna and P.~Z.~Skands: Comput. Phys. Commun., \textbf{178}, 852 (2008).
\bibitem{Bierlich} C.~Bierlich, G.~Gustafson and L.~Lonnblad: JHEP, \textbf{10}, 139 (2016).
\bibitem{CF} F.~Cooper and G.~Frye: Phys. Rev., \textbf{D10}, 186 (1974).
\bibitem{Andronic} A. Andronic \textit{et al}.: Nature, \textbf{561}, 321 (2018).
\bibitem{ALICE2} B.~Abelev \textit{et al}. (ALICE): Phys. Lett., \textbf{B734}, 409 (2014).
\bibitem{ALICE3} J.~Adam \textit{et al}. (ALICE): Phys. Lett., \textbf{B758}, 389 (2016).
\bibitem{ALICE4} B.~Abelev \textit{et al}. (ALICE): Phys.Rev., \textbf{C91}, 024609 (2015). %phi in PbPb
\bibitem{ALICE5} J.~Adam \textit{et al}. (ALICE): Eur.Phys.J, \textbf{C76}, 245 (2016). %phi in pPb
\bibitem{ALICE6} S.~Acharya \textit{et al}. (ALICE): arXiv:1807.11321. %(yield @Table.11自分で比を計算) %phi and proton in pp
\bibitem{ALICE7} B.~Abelev \textit{et al}. (ALICE): Phys.Rev., \textbf{C88}, 044910 (2013). %proton in PbPb
%\bibitem{jpsj} The abbreviation for the Journal of the Physical Society of Japan should be ``J. Phys. Soc. Jpn." in the reference list.
%\bibitem{instructions} More abbreviations of journal titles are listed in ``Instructions for Preparation of Manuscript", which is available at our Web site (http://jpsj.ipap.jp).
%\bibitem{format} F. Author, S. Author, and T. Author, Abbreviated journal title \textbf{volume in bold face}, initial page or article number (year of publication).
\end{thebibliography}
\end{document}